\documentclass{emulateapj}

\shorttitle{Prolate dark matter halo in the Milky Way}
\shortauthors{Banerjee \& Jog}

\begin{document}

\title{Progressively more prolate dark matter halo in the outer Galaxy as traced by flaring HI gas}

\author{Arunima Banerjee\altaffilmark{1} and Chanda J. Jog\altaffilmark{2}}
\affil{Department of Physics, Indian Institute of Science, Bangalore 560012, India}

\altaffiltext{1}{email: arunima$\_$banerjee@physics.iisc.ernet.in}
\altaffiltext{2}{email: cjjog@physics.iisc.ernet.in}

\begin{abstract}
A galactic disk in a spiral galaxy is generally believed to be embedded in an extended
dark matter halo, which dominates its dynamics in the outer parts. However, the shape of the halo is not clearly understood. Here we show that the dark matter halo in the Milky Way Galaxy is prolate in shape. Further, it is increasingly more prolate at larger radii, with the vertical-to-planar axis ratio monotonically increasing to 2.0 at 24 kpc. 
This is obtained by modeling the observed steeply flaring atomic hydrogen gas layer in the outer Galactic disk, where the gas is supported by pressure against the net gravitational field of the disk and the halo. The resulting prolate-shaped halo can explain several long-standing puzzles in galactic dynamics, for example, it permits long-lived warps thus explaining their ubiquitous nature. 
\end{abstract}


\section{INTRODUCTION}

Spiral galaxies are observed to have extended, nearly-flat rotation curves which indicate the existence of a dark matter halo. In fact, the radial distribution of the dark matter halo of a spiral galaxy is deduced from its observed rotation curve (Rubin 1983, Binney \& Tremaine 1987). However, the shape of the halo is not well-understood although it is expected to play a significant role in galaxy dynamics and evolution (Ryden 1990, Bekki \& Freeman 2002). Different observational tracers such as tidal streams or thickness of the interstellar gas layer so far have given only the average halo shape, which in 
most cases, is found to be either oblate or spherical. In general, this is also the shape assumed in theoretical studies for simplicity.

 The vertical thickness of the interstellar gas provides an additional constraint, since it allows one to trace the force normal to the plane and hence the shape of the dark matter halo on the galactic scale. But this approach has been applied only to a handful of cases so far (Becquaert \& Combes 1997, Olling \& Merrifield 2001, Narayan et al. 2005, Banerjee \& Jog 2008, Banerjee et al. 2010, O'Brien et al. 2010). The vertical thickness of the HI gas is observed to increase sharply with radius in the outer Galaxy, which 
cannot be explained by a dark matter spheroid of constant shape (Narayan et al. 2005), instead this requires a halo shape varying with radius (Kalberla et al. 2007).

In this paper, we model the observed steeply flaring HI gas to constrain the shape of the halo, and show that the halo in the outer Galaxy is prolate. Further, the axis ratio varies with radius such that the halo is progressively more prolate in the outer parts with a maximum vertical-to-planar axis ratio of 2.0 at around 8 disk scale lengths. In Section 2 we describe the formulation of the problem, the results are presented in Section 3. Section 4 contains a discussion of related points, and Section 5 summarizes our conclusions.

\section{FORMULATION OF PROBLEM}

\subsection{\emph{Model for Disk Vertical Structure}}

We have employed the multi-component model of the stars and gas of the galactic disk, in the field of the dark matter halo developed by Narayan et al. (2005).
For simplicity, each disk component is assumed to be isothermal.

The Poisson equation for an axisymmetric  galactic system in terms of the galactic cylindrical co-ordinates ($R$, $\phi$, $z$) is given by \\

$$\frac{{\partial}^2\Phi_{total}}{{\partial}z^2}
 = 4\pi G(\sum_{i=1}^{3} \rho_i + \rho_{h})
\eqno(1) $$ 

\noindent where $\rho_i$ with i = 1 to 3 denotes the mass density for each disk component, namely stars, HI and $H_2$, and $\rho_h$ denotes the same for the halo. Here $\Phi_{total}$ denotes the net potential due to the disk and the halo. For a flat or a gently-falling rotation curve, the radial term can be neglected as its contribution to the determination of the HI scale height is less than 10 percent as noted by earlier calculations (Narayan et al. 2005). 

The equation for hydrostatic equilibrium in the $z$ direction is given by (Rohlfs 1977)
$$ \frac{\partial}{{\partial}z}(\rho_{i}<(v_{z}^{2})_{i}>) + \rho_{i}\frac{{\partial}\Phi_{total}}{{\partial}z} = 0  \eqno(2) $$ \\
\noindent where $<(v_{z}^{2})_{i}>$ is the mean square random velocity along the $z$ direction for the component $i$. We further assume each component to be isothermal for simplicity, so that the velocity dispersion is constant with $z$. 

Eliminating $\Phi_{total}$ between eq. (1) and eq. (2), and assuming an isothermal case, we get
$$ <(v_{z}^{2})_{i}>  \frac{\partial}{{\partial}z}[\frac{1}{\rho_{i}}\frac{{\partial}\rho_{i}}{{\partial}z}] = -4\pi G(\sum_{i=1}^{3} \rho_i + \rho_{h}) \eqno(3)$$
which represents a set of three coupled, second-order differential equations, one for each component of the disk. 
These are solved together to obtain the vertical density distribution of each disk component.
This problem is solved in an iterative fashion, as an initial value problem, using the fourth order, Runge-Kutta method of integration. 
The two boundary conditions at the
mid-plane i.e $z = 0$ for each component are:
$$ \rho_i = (\rho_0)_i,  \qquad \frac{d\rho_i}{dz} = 0  \eqno(4) $$ \\
However, the modified mid-plane density $(\rho_0)_i $ for each
component is not known a priori. Instead the net surface
density $\Sigma_i(R)$,
given by twice the area under curve of
$\rho_i(z)$ versus z, is used as the second boundary condition,
since this is known from observations.
The HWHM (half width at half maximum) of the resulting model vertical distribution is used to define the vertical scale height.

Our earlier study of the Galaxy  based on this approach (Narayan et al. 2005), where a constant shape of the halo was used, showed that the observed flaring HI scale height distribution in the outer Galaxy can be explained by a spherical halo with the density falling faster than an isothermal case. This model, however, gave a total mass at the lower end of mass-range obtained by other techniques such as motions of satellites (Sackett 1999), which therefore questions its validity.

\subsection{\emph{Construction of the Dark Matter Halo Profile}}

Here we propose and try an alternative viable idea, namely, a prolate halo with a shape varying with radius. A prolate spheroid by definition has the vertical-to-planar axis ratio greater than 1, and it can be thought of as obtained by rotating an ellipse about its major axis. While a prolate halo fares better than a spherical halo of the same mass in explaining the flaring HI distribution, a single shape cannot explain the scale height data in the outer Galaxy over
a large range of radii $R$ = 9-24 kpc (Narayan et al. 2005, Kalberla et al. 2007). A halo that is progressively more prolate with radius is indicated  to explain the observed steep flaring. 

At any $R$, an isodensity contour for a spheroid is defined by a constant "$m$" where  $m^2  = {R'}^2 +  {z'}^2/{q_R}^2$ ; $R'$ and $z'$ are the co-ordinates of the points on the contour  and $q_R$ denotes the vertical-to-planar axis ratio of the spheroid at $R$. Here $R$ is the intercept on the mid-plane, that is, at $z = 0$. The mass of a shell bounded by two surfaces $m$ and $m + dm$  is a constant independent of the shape of the spheroid $q_{R}$ ( see eq. 2.74 of Binney \& Tremaine 1987 ), and this gives the following condition:
$$q_{R} \: \rho_{R} (q_{R}) \: = \: \rho_{R}(q=1) \eqno(5)$$								     
where $\rho_{R} (q_{R}) $ is the density along a prolate isodensity contour through $R$, and $ \rho_{R}(q=1)$ is the density along the corresponding spherical contour with radius $R$. Thus if the original spherical shell were instead taken to be a prolate spheroidal shell by construction, its density will be lower by a factor $q_{R}$. Thus the vertical force near the mid-plane, a crucial determinant of the HI scale height, will be lower and this could qualitatively explain the HI flaring that is observed. 

\subsection {\emph{Solution of Equations for a Prolate Halo}}

 We start with the Galactic mass model obtained by fitting the observed rotation curve (Mera et al. 1998) which gives the pseudo-isothermal spherical halo density distribution at a radius $r$ to be:

$$\rho(r) = \frac{\rho_0}{\large { 1+\frac{r^{2}}{{r_c}^{2}}\large}} \eqno(6) $$  
                                                                                                               
\noindent where $\rho_{0} = 0.035$ M$_{\odot}$ pc$^{-3}$ and $r_{c} = 5$ kpc are respectively the best-fit central density and the core radius for the halo.  This determines the mass within a spherical shell at each $r = R$. If the shape were taken as prolate instead, the density will change as per eq. (5) while the mass in the shell remains constant. One can ensure that the fit to the rotation curve still remains valid with deviation of only a few percent since the rotation velocity has a weak dependence on the shape of the halo mass distribution (Sackett \& Sparke 1990). In contrast, a small change in the halo shape of a given mass can have a striking effect on the vertical scale height. We exploit this idea to constrain the variation in the halo shape, $q_R$ , with the radius $R$. In a sense, each shell acts independently. We model $q_R$ as a simple second-order polynomial in $R$ i.e $q_R = 1. + {\alpha}_{1}(R-9) + {\alpha}_{2}(R-9)^2$ over a range $R$ = 9 - 24 kpc. Here $\alpha_{1},\alpha_{2}$ are free parameters. Combining this trial $q_R$ with eq. (5) and (6) gives the halo density, $\rho_{h}$, which is used as an input parameter in solving eq. (3).

The main aim of this paper is to investigate whether a prolate halo explains the observed HI flaring within the purview of the above mass model. We next calculate the HI scale height versus radius theoretically by solving eq. (3) numerically with the procedure as outlined in Section 2.1. The input parameters such as the stellar surface density, and stellar and gas velocity dispersions are taken to be the same as in our earlier work on the Galaxy (Narayan et al. 2005). The surface densities of HI and $H_2$ are taken from observations (Wouterloot et al. 1990).
The resulting HI scale height values obtained by solving eq. (3) for different trial are $q_R$ values are fit to the observed data (Wouterloot et al.1990) over the range  $R$ = 9-24 kpc. The halo is taken to be spherical at $R$ = 9 kpc.
Since the error  bars on the data are not available (Wouterloot et al. 1990), therefore we assume error bars of 5\%  and obtain  the best-fit values of the parameters ${\alpha}_{1}$,${\alpha}_{2}$ using the method of least squares. These are: $\alpha_1 = 0.020$ and $\alpha_{2} = 0.003$ and the range $\alpha_{1}$ = 0.010 - 0.080 and $\alpha_{2}$ = 0.001 - 0.004 gives reasonably good fit as well (Bevington 1969).

\section{RESULTS}

The plot of the resulting vertical scale height versus $R$, and the comparison with the observational data which were corrected for the Galactic warp (Wouterloot et al. 1990), is shown in Figure 1. The corresponding best-fit $q_R$ increases from a value of 1 corresponding to a spherical halo at $R = 9$ kpc to a prolate halo with $q_R  =$ 2.0 at $R = 24$ kpc. We stress that the $q_R$ value denoting the halo shape as obtained here is a local property, and our approach has allowed us to obtain a radial variation in it. In contrast, studies involving tidal streams (Ibata et al. 2001, Law \& Majewski 2010) and other observational tracers  assumed the halo shape to be constant, and found it to be typically spherical or oblate ($q_R$ $<$ 1) (Sackett 1999). There are a few exceptions involving studies of microlensing (Holder \& Widrow 1996) and stellar streams (Helmi 2004) which yield a prolate halo, but again with a global, constant shape $q_R$ $ < $ 1.5  within $R = 60$ kpc. 

Figure 2 gives the isodensity contours (on the $R$-$z$ plane) denoting this increasingly prolate halo in the outer Galaxy, and the inset shows the resulting $q_R$ versus $R$. We emphasize that our study traces the halo shape variation over a large radial distance upto 24 kpc, that is 8 disk scale lengths (with the disk scale length being 3.2 kpc as in the Mera et al. (1998) model) or twice the optical radius of the Galaxy. Surprisingly, a prolate halo with a maximum $q_R$ = 2.0 at $R$ = 24 kpc  is sufficient to explain the steep rise in the gas thickness by a factor of 6 seen over a radial range of 9-24 kpc. This is because the density in a prolate shell at a given $R$ is inversely proportional to $q_R$ (see eq. 5), while the vertical force due to gravity and hence the resulting scale height depends in a nonlinear way on the density near the midplane (Banerjee \& Jog 2007). Thus a small decrease in the midplane density  is sufficient to explain the observed sharp increase in scale heights.

\section{DISCUSSION}

\noindent {\bf \emph{1. Dependence on gas velocity dispersion:}} The value of gas dispersion plays a crucial role in these calculations, but it is not easy to measure and depends on the fraction of HI in the different phases.
In the outer Galaxy the ratio of the 21 cm emission and absorption is shown to be remarkably constant upto $R$=25 kpc
which implies a constant ratio of the warm and cold phases of HI  (Dickey et al. 2009).   
However, the pressure support for the gas is mainly from non-thermal or turbulent motions which dominate the thermal velocity dispersion, and Dickey et al. (2009) do not give the non-thermal velocitites.
Hence here we use the measured dispersion at the solar location of 9 ${kms}^{-1}$ (Malhotra 1995). This is expected to decrease at larger radii and saturate to 7 ${kms}^{-1}$ (see Narayan et al. 2005). These values were used as input parameters in our calculations. These values are in good agreement with the radial variation seen in external galaxies (Kamphuis 1993, Tamburro et al. 2009). 

\noindent {\bf \emph{2. Mass of the prolate dark matter halo:}} In this paper we start with the screened isothermal 
mass model of the Galaxy as given by Mera et al. (1998), which has a mass of 10$^{12}$ M$_{\odot}$ within $R < $ 100 kpc.
 Due to the construction adopted in this paper, the mass in each shell is conserved as an initially 
spherical shell is stretched into a prolate shell (eq. [5]). Thus the total mass of 
the halo is conserved. Hence the total mass of the prolate halo in our model is 
the same as in the Mera et al. mass model.   We note that in our model we have only dealt with the region of $R < 24$ kpc.

This total mass is in a good agreement with the value obtained using different observational tracers such as motions of satellites, and high velocity stars (Sackett 1999, also see Gnedin et al. 2010, McMillan 2011 for more recent estimates). 
Thus the current model overcomes the problem with the earlier model (Narayan et al. 2005) which gave a total halo mass within 100 kpc that was three times smaller than the above value. Recall that this was the motivation 
for trying the variation in the halo shape with radius (Section 2).
Thus our model explains the observed HI scale heights and also gives a total mass in agreement with typical values in the literature.

\noindent {\bf \emph{3. Asymmetry in gas scale heights:}} The scale height distribution is known to be asymmetric in the two halves of the Galaxy (Levine et al. 2006, Kalberla et al. 2007)  being higher by a factor of about two in the northern galactic hemisphere than in the south. We have only modeled the northern data (Wouterloot et al. 1990) here 
for simplicity to check if our proposed model of a prolate dark matter halo with a radially-dependent shape could explain the steeply increasing HI scale height data. To model the southern region, we would need the observed HI surface density 
values as input parameters.
Levine et al. (2006) measure the HI scale height for both north and south, but only the net value of the gas surface density, averaged over the north and south is given. 

 Recently, Kalberla \& Kerp (2010) have given observed HI surface density values separately for the north and south. 
Applying our model to the southern data alone, we find that the best-fit $q_R$ values are significantly lower. 
A prolate halo with a radially-increasing axis ratio is still preferred over an oblate or spherical case, 
but with  smaller values of $\alpha_1$ = 0.009 and $\alpha_2$ = 0.0
(compared to 0.02 and 0.003 respectively for the northern data, Section 2.3). These give a maximum axis ratio of 1.14 at 
$R$=24 kpc as compared to $q_R$ = 2.0 obtained for the north. This low value of $q_R$ is sufficient to explain the data 
in the south, since the scale heights are lower and the rate of flaring beyond 18 kpc is much less steep in the south than in the north.
 Thus the dark matter halo of the Galaxy has a prolate shape but its value is different in the two hemispheres, being more prolate in the north. A more realistic future treatment should consider simultaneous modeling of north and south HI scale height data.

There is a limitation in applying this model farther out in the plane since the gas dispersion values are not well-known.
With this caveat in mind, if we model the data the range $9-40$ kpc (Kalberla \& Kerp 2010) by extrapolating the above input parameters, the best fit gives a maximum $q_R$ =  4.5  at $R = 40$ kpc. We have not modeled the inner Galaxy 
($ R <$ 9 kpc) here,  since the halo contribution is shown to be small in this region (Narayan \& Jog 2002), hence the scale heights cannot be used to constrain the halo shape.

\noindent {\bf \emph{4. Comparison with shapes from cosmological simulations:}}
 Interestingly, the prolate-shaped dark matter halo obtained here by modeling the HI vertical scale height data agrees well with the general trends seen in cosmological simulations. The latter give a range of halo shapes with a preference for a prolate shape (Bailin \& Steinmetz 2005, Bett et al. 2007) but with a lower vertical-to-planar axis ratio. However, these are measured at scales of about 100 kpc (Bailin \& Steinmetz 2005)
 which are much larger than that of a galactic disk, hence the quantitative comparison is not meaningful. The predictions of our model can be checked and confirmed by the signatures of other local tracers in future observational studies such as GAIA.

\noindent {\bf \emph{5. Dynamical implications of a prolate halo:}}
A prolate-shaped dark matter halo as obtained here has important implications for many classic problems in galactic dynamics. For example, a prolate halo would cause a lower differential precession and hence can support long-lasting warps (Ideta et al. 2000), this can explain why warps are commonly seen. Second, a prolate halo can naturally explain (Helmi 2004) the fact that the satellites are seen to be limited to the polar plane normal to the galactic plane, as observed in many galaxies, known as the Holmberg effect. The dynamical implications of a prolate halo, in particular
one which is increasingly more prolate with radius, deserve a detailed study.

\section{CONCLUSIONS}
By modeling the flaring HI gas distribution, we have shown that the shape
of the dark matter halo is prolate in the outer Galaxy over a radial range of $R = 9-24$ kpc, where it can have possible observable dynamical consequences. We treat the halo as a set of spheroidal shells which have progressively more prolate shapes with increasing radii. Thus, we obtain the local shape of the halo and show that 
the maximum vertical-to-planar axis ratio is 2.0 at $R$ = 24 kpc. These results are in contrast with most of the earlier work involving various observational tracers, which gave either a
spherical or a flattened, oblate halo with a constant shape. However, our results agree with the trends from cosmological simulations which tend to favor a prolate halo. A prolate halo, in particular one which is increasingly more prolate at larger radii, has important implications for galaxy dynamics and evolution, which need to be studied further.

\bigskip 

\noindent {\bf Acknowledgments}

\medskip

We would like to thank the anonymous referee for insightful comments on the paper.

\clearpage

\begin{figure}
\epsscale{.80}
\plotone{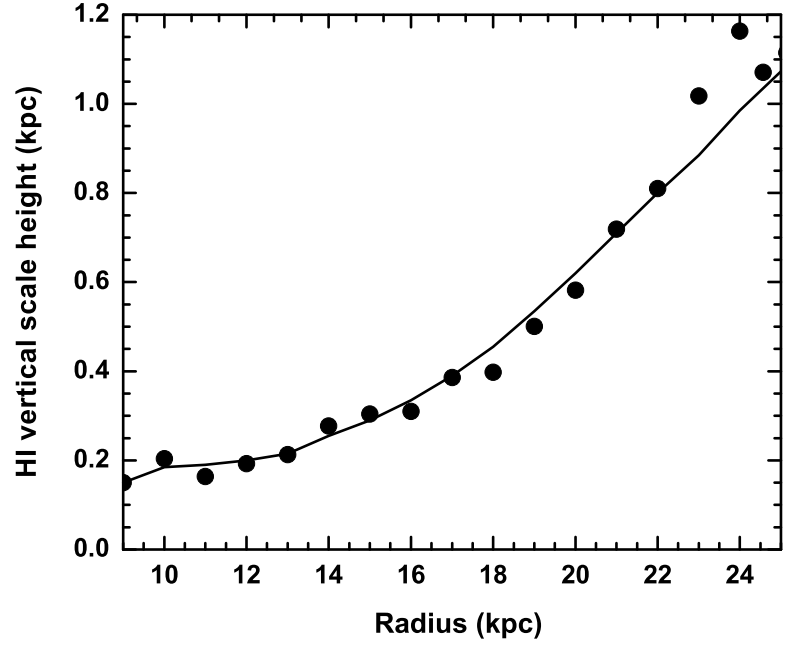}
\caption{The calculated vertical scale height for the atomic hydrogen gas, HI, (solid line) and the observational values (Wouterloot et al. 1990) (filled circles) versus Galactocentric radius $R$. The theoretical curve is the best-fit case, and corresponds to a dark matter halo which is progressively more prolate with radius.
In the radial range studied, the halo is found to be most prolate with the vertical-to-planar axis ratio, $q_R$ = 2.0 at $R$ = 24 kpc. \label{fig1}}
\end{figure}

\clearpage

\begin{figure}
\epsscale{.80}
\plotone{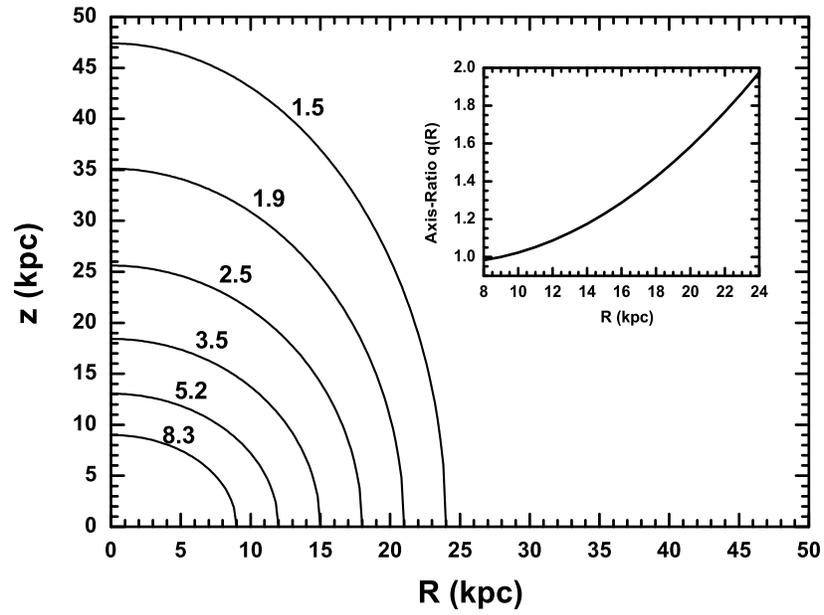}
\caption{The resulting best-fit prolate-shaped isodensity contours of the dark matter halo on the $R$-$z$ plane, with vertical-to-planar axis ratio  increasing with radius as depicted in the inset. This clearly brings out the increasingly prolate geometry of the halo shape. The densities of the successive contours moving radially outwards are 8.3, 5.2, 3.5, 2.5, 1.9, and 1.5 in units of ${10}^{-3}$ M$_{\odot}$ pc$^{-3}$. \label{fig2}}
\end{figure}

\end{document}